# Femtogram Doubly Clamped Nanomechanical Resonators Embedded in a High-*Q* Two-Dimensional Photonic Crystal Nanocavity

Xiankai Sun,[1] Jiangjun Zheng,[2] Menno Poot,[1] Chee Wei Wong,[2] and Hong X. Tang[1,*]

[1]*Department of Electrical Engineering, Yale University, New Haven, Connecticut 06511, United States*
[2]*Optical Nanostructures Laboratory, Columbia University, New York, New York 10027, United States*
*Corresponding author: hong.tang@yale.edu

ABSTRACT: We demonstrate a new optomechanical device system which allows highly efficient transduction of femtogram nanobeam resonators. Doubly clamped nanomechanical resonators with mass as small as 25 fg are embedded in a high-finesse two-dimensional photonic crystal nanocavity. Optical transduction of the fundamental flexural mode around 1 GHz was performed at room temperature and ambient conditions, with an observed displacement sensitivity of 0.94 fm/Hz$^{1/2}$. Comparison of measurements from symmetric and asymmetric double-beam devices reveals hybridization of the mechanical modes where the structural symmetry is shown to be the key to obtain a high mechanical quality factor. Our novel configuration opens the way for a new category of "NEMS-in-cavity" devices based on optomechanical interaction at the nanoscale.

Flexural nanomechanical resonators have consistently set records for sensitive measurements of mass, force, and displacement [1-9]. These resonators with an ultrasmall mass are of particular interest because they inherently operate at the ultrahigh frequency of the fundamental mechanical mode, which is advantageous for developing high-speed sensors with ultimate sensitivity. At the same time, achieving a high mechanical quality (*Q*) factor, particularly in ambient setting, is equally important because the transducer sensitivity and the coherence time of the mechanical vibration are proportional to the *Q* factor. On the other hand, although recent developments in one-dimensional (1D) photonic crystals (PhC) have enabled optomechanical transduction at high frequencies [10,11], the resonating mode is often post-selected among many acoustic modes simultaneously supported in an optimized photonic structure. The coupling between a single optical cavity mode and commonly utilized flexural mechanical vibrational modes thus still remains challenging. Additionally, the demonstrated high-frequency modes are based on high stiffness rather than small motional mass. Furthermore, the poor heat dissipation of the 1D PhC limits the highest operating power of the cavity [12,13]. Therefore, an optomechanical system with a well-defined, ultrasmall-mass, high-*Q* mechanical resonator that overlaps strongly with a high-*Q* optical cavity with large power handling capability is highly desirable for extremely sensitive measurements involving mass, force, and displacement. Here we implement the first realization of the idea of a "NEMS-in-cavity," by embedding a femtogram doubly clamped nanomechanical double-beam resonator in a finely tuned two-dimensional (2D) PhC nanocavity. Conceptually this "nanobeam-in-cavity" configuration is analogous to cavity quantum electrodynamics (cQED) system realized by embedding a single emitter (e.g., an atom, molecule, or quantum dot) in a high-$Q$ optical cavity. Here, we carve out a nanomechanical resonator with well-defined vibration modes (in place of the emitter of cQED) within a high-$Q$ photonic nanocavity.

The simplest approach for realizing a "NEMS-in-cavity" would be directly enclosing a tiny nanomechanical resonator within a well-established PhC nanocavity such as the L3 cavity [14]. However, due to the strong perturbation to the cavity mode by the embedded nanobeams, the PhC nanocavity needs to be thoroughly redesigned and engineered to ensure maximum confinement of the cavity mode. Careful optimization results in a single optical mode with a high intrinsic optical $Q$ of 19,500. Enabled by the well-defined geometry and femtogram mass of the mechanical resonator, optical transduction of the nanobeam's single fundamental mechanical flexural mode around 1 GHz is demonstrated, with a mechanical $Q$ of 1230 in vacuum and 580 in air. Further investigation on symmetric and asymmetric double-beam devices reveals hybridization of the mechanical modes and shows that the perfect symmetry of the double beams is crucial for obtaining a high mechanical $Q$.

Beyond demonstrating sub-fm/Hz$^{1/2}$ sensitivity at the near GHz frequency, our "NEMS-in-cavity" approach has several additional advantages. First, not only is the optical modal volume minimized, but also the mechanical mode volume is reduced well below $(\lambda_0)^3$, where $\lambda_0$ is the free-space wavelength of the optical mode. In our devices, the effective volume of the mechanical mode is about 10% of that of the optical cavity mode, yielding large optomechanical coupling rates. Furthermore, by using well-studied doubly clamped beams, engineering nanomechanical resonators with different geometry parameters is straightforward and largely independent from the optical cavity design, thus allowing for the wide use in various precision sensing and metrology applications. Also, the use of 2D PhC as the platform, as opposed to 1D PhC nanocavities [10,11,15,16], facilitates the heat dissipation and thus remarkably improves the power handling capability of the devices. Finally, by using a CMOS-compatible fabrication process, our optomechanical structures are laid out along with integrated grating couplers and waveguides on an all-integrated Si photonics platform [17]. The integrated circuit approach provides an efficient framework for characterizing such optomechanical devices because it provides measurement stability and allows for the critical coupling condition between the PhC waveguides and the PhC nanocavity to be reliably achieved through design and lithographic patterning. The vertical symmetry of this in-plane coupling scheme also helps to suppress optomechanical coupling of low-frequency PhC membrane modes [18]. This work lays the foundations for a new category of "NEMS-in-cavity" devices based on optomechanical interaction at the nanoscale.

Various approaches to obtain high-$Q$ two-dimensional Si nanocavities could be followed, including neighbor-hole-shifted three-lattice-point (L3) cavities [14], double-heterostructure cavities [19], and width-modulated line-defect cavities [20,21]. Here, a variant of the L3 cavity is employed because this configuration naturally fits in our "beam-in-cavity" concept for beams of submicron length. As shown in Figure 1a and b, we create a nanomechanical double-beam resonator inside a PhC L3 cavity. This is done by placing three parallel slots, separated from each other by the beam width $W_{beam}$ and with their lengths being the beam length $L_{beam}$, thus forming a "nanobeam-in-cavity" optomechanical system. In principle, one could also use single-beam configuration. However, a double-beam cavity is more favorable due to its overall symmetry, which is important to reduce radiation losses and achieve a high mechanical $Q$. To demonstrate the beam's fundamental flexural mode vibrating at around 1 GHz (a frequency well in the UHF range and yet measurable with our experimental setup), the beam length and width are chosen to be $L_{beam}$ = 785 nm and $W_{beam}$ = 80 nm or 90 nm. The corresponding slot width is set to be $W_{slot}$ = 60 or 53 nm, where the area occupied by the mechanical resonator is kept the same to minimize the variation of the optical mode. The effective mass is 25.6 fg or 27.8 fg for $W_{beam}$ = 80 nm or 90 nm, respectively. Note that the nanobeams in our scheme can be designed *a priori* and is mostly independent of the cavity, which provides great freedom in engineering the appropriate mechanical resonator for specific applications, in stark contrast to previously demonstrated "cavity-in-beam" and "beam-cavity" structures where the mechanical resonator cannot be clearly defined, and its geometry and the associated cavity mode have to be adjusted iteratively [10,11,22].

The inclusion of a nanomechanical resonator induces a strong perturbation to the original L3 cavity mode, and thus the cavity had to be redesigned to recover the high-$Q$ mode. The creation of the three slots (by replacing the refractive index of Si with that of the air) reduces the effective index of the cavity and shifts the cavity band up into the slab mode continuum, making the cavity mode a leaky mode with a low $Q$ [23]. To counteract the effects of refractive index reduction, we increased the width of the cavity row by a factor of 1.35 (i.e., a vertical hole-to-hole distance of $1.35\sqrt{3}a$, where $a$ is the PhC lattice constant). This strategy pulls the cavity mode back into the center of the bandgap where the mode confinement is maximal, and this minimizes the in-plane optical loss. Figure 1c shows the TE-like band diagram of such a W1.35 waveguide (with a W1 waveguide as comparison) with the three embedded slots (extended infinitely in the $x$ direction), as computed with MPB [24], a fully-vectorial eigenmode solver of Maxwell's equations with periodic boundary conditions. The lattice constant $a$ = 430 nm, hole radius $r$ = 0.279$a$ = 120 nm, thickness $t$ = 220 nm, slot width $W_{slot}$ = 60 nm, and beam width $W_{beam}$ = 80 nm. At the middle of the bandgap sits a single band of the W1.35 waveguide (violet solid line), with its mode concentrated mostly inside the slots (see the inset of Figure 1c), which ensures good transverse modal confinement. By comparison, the corresponding band of the W1 waveguide (violet dash line) is buried in the upper continuum of the PhC slab modes.

We then optimize the PhC nanocavity with slots of the prescribed length $L_{beam}$. To achieve good longitudinal modal confinement, the holes to the sides of the slots in the cavity row are enlarged accordingly to a radius of 160 nm, thus maintaining the same filling ratio of the other part of the PhC membrane. The cavity resonant wavelength $\lambda_0$ and quality factor $Q$ were simulated with MEEP [25] by a three-dimensional finite-difference time-domain (FDTD) method. The highest optical $Q$ factor is obtained by adjusting the positions of ten holes surrounding the cavity, with six holes in the cavity row (three on each side) shifted in the $x$ direction (outwardly by $S_{x1}$, $S_{x2}$, and $S_{x3}$, respectively) and four in the cavity neighboring rows (two in each row) shifted in the $y$ direction (outwardly by $S_y$) (see Figure 1a). With preset beam parameters $W_{slot}$ = 60 nm and $W_{beam}$ = 80 nm, the highest cavity $Q$ was found for a structure with parameters $L_{beam}$ = 1.8$a$, $S_{x1}$ = −0.18$a$, $S_{x2}$ = −0.06$a$, $S_{x3}$ = 0.22$a$, and $S_y$ = −0.15$a$. This mode has a resonant wavelength $\lambda_0$ = 1541.7 nm, a theoretical quality factor $Q$ = 19,500, and an effective modal volume $V_o$ = 0.022 $(\lambda_0)^3$. The geometry used in the simulation does not include the effects from fabrication imperfections. The electric field component $E_y$ is shown in Figure 1d, and its two-dimensional Fourier transform (Figure 1e) shows negligible components in the leaky region, indicative of a high-$Q$ mode.

Next, to accurately simulate the mechanical modes the above structure was imported into COMSOL [26], an eigenmode solver based on three-dimensional finite-element method (FEM). Their mechanical frequency ($f_m = \Omega_m/2\pi$) and modal displacement profile $\mathbf{U}(x, y, z)$ were directly obtained from the FEM simulation. The effective mass is calculated with the definition $m_{eff} = \int dV \rho |\mathbf{U}|^2/\max(|\mathbf{U}|^2)$, where $\rho$ is the density of the material. The optomechanical coupling strength $g_{om}$, defined as $d\omega_c/du$ ($\omega_c$ is the cavity mode frequency and $u$ denotes the displacement of the mechanical mode), can be determined by the optical cavity field and the mechanical displacement profile using a perturbation theory for optomechanical systems [11,27]. Two modes pertaining to the beam's fundamental mode were found, corresponding to the differential and common motions of the double beams. The differential mode (Figure 1f) with frequency $\Omega_m/2\pi$ = 964.9 MHz and effective mass 25.6 fg has a strong optomechanical coupling $g_{om}/2\pi$ = 10.8 GHz/nm. The common mode (Figure 1g) with frequency $\Omega_m/2\pi$ = 976.0 MHz and effective mass 29.5 fg has a weak optomechanical coupling $g_{om}/2\pi$ = 0.31 GHz/nm.

The devices were fabricated from standard silicon-on-insulator (SOI) substrates (Soitec UNIBOND™), with a 220-nm Si layer on 3-μm buried oxide. The entire structure was patterned by high-resolution electron-beam lithography (Vistec EBPG 5000+) of a positive-tone resist. Then the pattern was transferred by $Cl_2$-based inductively-coupled plasma reactive ion etching (Oxford PlasmaLab System 100) to the Si device layer. Finally, the PhC membrane was released from the substrate by photolithography and subsequent wet etching in a buffered oxide etchant.

Figure 2a shows a typical device which includes a pair of grating couplers for vertically coupling light onto and out of the chip, strip waveguides for routing light into PhC structure, and PhC waveguides for coupling light into and out of the cavity. The PhC membrane terminates with a termination parameter τ = 0 as defined in Ref. 28 to facilitate a low-reflective interface between the strip waveguide and the PhC waveguide. The positions of the ends of PhC waveguides relative to the cavity were determined both numerically and experimentally to ensure the maximal on-resonance transmission of the cavity mode [29].

First, we characterized the fabricated devices optically. As shown in the experimental setup in Figure 2a, the device chip was placed in a vacuum chamber, which was pumped below 0.1 mbar to minimize the gas damping effect. The light of a C-band tunable diode laser (Santec TSL-210) was attenuated and its polarization was adjusted before sent to the devices. The light enters the chip via the first grating coupler and is routed toward the cavity. Light passing through the cavity is collected into an optical fiber via a second grating coupler. This transmitted signal was split by a 99/1 fiber coupler. 1% of the split light was used for monitoring the transmission level and recording the transmission spectrum with a kHz photodetector (New Focus model 2011). The remaining 99% of the split light was sent through a fiber preamplifier (Pritel FA-20) before reaching a GHz photodetector (New Focus model 1611). The detected signal was then sent to an electrical spectrum analyzer (Hewlett Packard 4396A) to measure the radio-frequency (RF) power spectral density containing the mechanical signal.

Depending on the specific parameters of PhC lattice and beam structures, the fabricated devices have a single optical resonance between 1520–1570 nm. Figure 2b shows a typical transmission spectrum of a device, displaying a single optical resonance at 1548.49 nm. A Lorentzian fit of its narrowband spectrum reveals a loaded optical $Q$ factor of 10,000, leading to a similarly high finesse for this low-order cavity mode. The good agreement between simulated and measured resonant wavelength and optical $Q$ factor indicates our mature control of device fabrication. The normalized on-resonance transmission is $1.23 \times 10^{-3}$. Taking into account a 16 dB coupling loss typically introduced by the pair of grating couplers, the insertion loss between the input and output strip waveguide is estimated to be 13 dB, which consists of the modal mismatch loss at the joints of strip waveguide to PhC waveguide, the propagation loss inside the PhC waveguides, and the tunneling loss between the ends of PhC waveguide and the cavity.

We measured the double beam's mechanical modes by setting the laser wavelength at the maximum slope of the optical resonance and recording the noise spectrum of the optical transmission. The nanobeam's thermal vibration causes phase variations of the optical cavity mode, which induces a resonance shift at the frequencies of the mechanical modes and results in an intensity modulation at a wavelength near the (fast-shifting) resonance. Therefore the noise spectrum of the optical transmission contains the signature of the nanobeam's vibrational modes [30]. Figure 2c and d show the RF spectrum of the mechanical mode of a $W_{\text{beam}}$ = 80 nm device. In the entire spectrum, a single mechanical mode at 903.6 MHz is observed, which is considered to be the differential mode because of its dominantly stronger optomechanical coupling to the optical cavity mode. The common mode is not observed due to its weak coupling (i.e., small $g_{om}$) to the optical mode and also its higher mechanical damping compared to the differential mode, as will be discussed later. The measured mechanical $Q$ values are 1230 in vacuum and 580 in air. These $Q$ factors are comparable with those from other nanomechanical resonators with similar dimensions [10,31]. Among various designs, the highest measured frequency is from a $W_{beam}$ = 90 nm device, also exhibiting a single mechanical mode, at frequency 1.081 GHz (Figure 3b). All measurements were performed at low laser intensity to ensure that optomechanical amplification or damping from dynamic back-action is negligible. This is confirmed by the same mechanical $Q$ values for blue and red detuning of the laser from the optical resonance [30].

The displacement noise power spectral density (PSD) of the thermomechanical motion not only gives the resonant frequency and $Q$ factor, but also provides a reliable way to calibrate the sensitivity of the measurement system. In the experiment, the relation between the displacement and the photodetector voltage is *a-priori* unknown. However, the area under the peak in the RF PSD corresponds to the Brownian motion of the resonator and thus provides a way to calibrate this transduction factor. Focusing on the $W_{beam}$ = 80 nm device of Figure 2d, its calculated effective mass (25.6 fg) and measured frequency (903.6 MHz) yield an elastic constant $k$ = 825 N/m. This corresponds to a root-mean-square displacement amplitude of $u_{rms} = (k_B T/k)^{1/2} \approx 2.24$ pm at room temperature $T$ = 300 K. The displacement sensitivity is set by the noise floor of the RF spectrum and it depends on the laser power and the transmission responsivity (slope of the transmission vs. wavelength). In Figure 2d, for the measurement in air with 2.5 mW optical power in the feeding waveguide, a displacement sensitivity of 0.94 fm/Hz$^{1/2}$ is achieved, which is a factor of 77 above the standard quantum limit. This value is among the highest that have been demonstrated around GHz frequencies and is at the same order of magnitude of other sensitive nano-optomechanical systems at much lower frequencies [7-9,32]. Comparison with the measurement in vacuum shows a higher input power indeed helps in achieving a better displacement sensitivity. Note that the input optical power of 2.5 mW in the feeding waveguide represents at least an order of magnitude improvement of the power handling capability of our device compared to that demonstrated in 1D PhC [10].

The measured mechanical $Q$ of ten devices with the above symmetric-design double beams has a broad distribution varying from 610 to 1230 (see, e.g., Figure 3a and b). In order to explain this behavior, we employ a model based on two coupled mechanical oscillators with effective masses $m_1$, $m_2$ and elastic constants $k_1$, $k_2$, corresponding to the two individual nanomechanical beams of our system. The equations of motion for this coupled oscillator system are [33]

$$m_1\ddot{u}_1 + m_1\gamma_a\dot{u}_1 + \gamma_c(m_1\dot{u}_1 + m_2\dot{u}_2) + k_1u_1 + k_c(u_1 - u_2) = F_1, \qquad (1)$$

$$m_2\ddot{u}_2 + m_2\gamma_a\dot{u}_2 + \gamma_c(m_1\dot{u}_1 + m_2\dot{u}_2) + k_2u_2 + k_c(u_2 - u_1) = F_2. \qquad (2)$$

In the above equations, $\gamma_a$ denotes the coupling-independent dissipation, such as material loss, air damping, and the coupling-independent portion of clamping loss. $k_c$ is the coupling coefficient of the two oscillators. $F_1$ and $F_2$ are the thermal Langevin forces acting on the oscillators. The $\gamma_c$ terms denote the difference in clamping loss for the collective motion of the coupled oscillator system. Its origin can be understood as follows: motion of the individual beams creates a stress profile in the surrounding PhC membrane, which is different for in-phase and anti-phase motion. This implies that the dissipation channel due to emitting phonons into the substrate is different in the two cases [34].

The eigenmodes of the coupled equations (1) and (2) are solved to obtain the modal frequency and modal damping rate. Due to the dissipative ($\gamma_c$) and dispersive ($k_c$) coupling, the original individual modes are hybridized, forming an in-phase ($u_{IP}$) and an anti-phase ($u_{AP}$) coupled mode (details provided in the Supporting Information). In the case that two nonidentical but similar oscillators are coupled with a small asymmetry parameter $\delta$ defined by $(\Omega_1 - \Omega_2)/2$ where $\Omega_{1,2} = \sqrt{(k_{1,2}+k_c)/m_{1,2}}$ is the angular frequency of the individual beam, the relative weights of the individual modes are almost equal, thus producing an in-phase common mode $u_{IP} \approx (u_1 + u_2)/2$ with damping rate $\gamma_a + 2\gamma_c\left[1-\delta^2/(\gamma_c^2+v_c^2)\right]$ and an anti-phased differential mode $u_{AP} \approx (u_1 - u_2)/2$ with damping rate $\gamma_a + 2\gamma_c\delta^2/(\gamma_c^2+v_c^2)$, where $v_c$ is a frequency parameter defined by $2k_c/\left[(\Omega_1+\Omega_2)(m_1 m_2)^{1/2}\right]$ which characterizes the dispersive coupling between the two oscillators. When the two oscillators are identical, i.e., $\delta = 0$, the common mode has its highest damping rate $\gamma_a + 2\gamma_c$ and the differential mode has its lowest damping rate $\gamma_a$. On the other hand, if the two oscillators are quite different from each other, the coupled modes reduce to the individual uncoupled modes, i.e., $u_{IP} \approx u_1$ and $u_{AP} \approx u_2$. Their modal frequencies almost maintain the original uncoupled values, and their damping rates are almost equal, given by $\gamma_a + \gamma_c$.

With this model, it becomes clear that the only mode observed in the symmetric-design double-beam devices is the differential mode, which compared with the common mode not only has a higher optomechanical coupling $g_{om}$ but also has a lower damping rate (or, equivalently, higher mechanical $Q$). The broad distribution of the measured $Q$ values can be attributed to the asymmetry ($\delta \neq 0$) of the double beams, which is introduced by the small and uncontrollable fabrication imperfections. When $\gamma_a$ is much less than $\gamma_c$, the damping rate of the differential mode is dominated by the $\gamma_c$ term, which is directly proportional to $\delta^2$. Therefore, the distribution of such asymmetry ($\delta$) is translated into the measured $Q$ values of the symmetric-design devices, varying from 1230 to 610 (Figure 3a and b).

The origin of different damping rates for the differential and common modes lies in their inherent motional behavior. As explained above, the damping rates are closely related to the radiating elastic strain field. The FEM simulation (in the Supporting Information) indicates that the common mode vibration acts as a dipole source and induces a transverse wave in the PhC membrane radiating away from the oscillator. However, due to the prohibited dipole radiation by the destructive interference, the differential mode acts as a quadruple source and induces a longitudinal wave radiating away from the oscillator, thus experiences much less mechanical dissipation because of the much weaker radiation field.

In order to estimate the relative strength of different damping mechanisms, we also fabricated asymmetric-design double-beam devices in which the center slot is shifted 3 nm transversely, causing a 6-nm difference between the widths of the double beams. As confirmed both numerically and experimentally, this shift does not cause any observable variation in the optical $Q$ from that of the symmetric-design double-beam devices. According to the beam theory [35] and numerical simulation, this beam width difference results in a frequency difference $2|\delta| \approx 70$ MHz for $W_{beam} = 80$ nm devices. This asymmetry brings the coupled oscillator system into the weakly coupled regime, and thus the two eigenmodes will have their modal frequencies close to the individual uncoupled values, with approximately equal damping rates $\gamma_a + \gamma_c$. Additionally, such an asymmetry also results in approximately equal optomechanical coupling for both eigenmodes ($g_{om}/2\pi \approx 5.5$ GHz/nm, i.e., half of the sum of the coupling rates of the common and differential motion). The RF spectra of such intentionally asymmetric devices are displayed for a $W_{beam} = 77$ nm/83 nm device (Figure 3c) and for a $W_{beam} = 87$ nm/93 nm device (Figure 3d), where the modes possess their mechanical $Q$'s of around 55. Assuming the highest $Q$ ever achieved from the symmetric-design devices is from one having a perfect symmetry ($\delta = 0$), it is thus straightforward to establish a relation between the experimental $Q$ and the damping rates

$$\frac{\gamma_a}{\gamma_a + \gamma_c} = \frac{Q_{asym}}{Q_{sym}} = \frac{55}{1230}, \qquad (3)$$

which leads to a ratio of 21.4 between the two damping rates $\gamma_c$ and $\gamma_a$. This experimentally determined ratio of $\gamma_c/\gamma_a$ confirms in turn that the mechanical $Q$ of such double-beam devices is actually limited by the $\gamma_c$ term, the clamping loss due to the collective motion of the double beams, and that the perfect symmetry of the double beams is the key to achieve a high $Q$ of a single mechanical mode, i.e., the differential mode.

In summary, we have demonstrated for the first time a "nanobeam-in-cavity" optomechanical system on a Si integrated photonics platform, which consists of a femtogram doubly clamped nanomechanical resonator embedded in an engineered high-$Q$ two-dimensional photonic crystal nanocavity. By using the well-studied doubly clamped beams as the nanomechanical resonator, the mechanical design is easy and independent from the optical cavity design, thus allowing for versatile geometries of the mechanical resonator for various applications. As a side-clamped mechanical beam oscillator with length $L$, width $W$, and thickness $t$ has its mass $m \propto LW$ and elastic constant $k \propto W^3/L^3$ when the vibration is along the width dimension, combinations of $L$ and $W$ can lead to a wide range of $m$ and $k$. For example, in weak force measurement such as magnetic resonance force microscopy [36] where beams with very low elastic constant are required, we may design the beams to have small width or long length, or even use a single-clamp geometry. Another example lies in the mass sensing application where a small mass is desirable, because the smallest resolvable mass is $\Delta m = 2m_{\text{eff}} (\Delta f_m/f_m)$ for a given frequency resolution $\Delta f_m$. Another bonus of reducing the resonator mass would be an enhanced optical $Q$ (after optimization), resulting from a less perturbed cavity. Actually structures of 60-nm beams achieve a theoretical optical $Q$ above 50,000 from our numerical simulation, an enhancement by a factor of 2.5 over the current generation of 80-nm beams. This optical $Q$ enhancement would lead to higher measurement sensitivity because of the stronger signal readout. The analysis from the results of symmetric and asymmetric double-beam devices reveals hybridization of the mechanical modes and shows that the perfect symmetry of the double beams is the key to obtain a high mechanical $Q$. Future work will be aimed at further enhancing the optical and mechanical $Q$ values to make such devices highly useful in cavity quantum optomechanics.

**Supporting Information.** Detailed derivation of the solutions to the coupled equations (1) and (2). Modal frequency and damping rate of the eigenmodes of a coupled oscillator system under different conditions. FEM simulation of the stress and displacement fields of the common and differential mode of the double beams in PhC cavity.

ACKNOWLEDGMENT
We acknowledge funding from DARPA/MTO ORCHID program through a grant from the Air Force Office of Scientific Research (AFOSR). H.X.T. acknowledges support from a National Science Foundation CAREER award and Packard Fellowship in Science and Engineering. J.Z. and C.W.W. acknowledge support from National Science Foundation CAREER and IGERT awards. M.P. acknowledges a Rubicon fellowship from the Netherlands Organization for Scientific Research (NWO)/Marie Curie Cofund Action. The authors thank Michael Power and Dr. Michael Rooks for assistance in device fabrication.

FIGURES

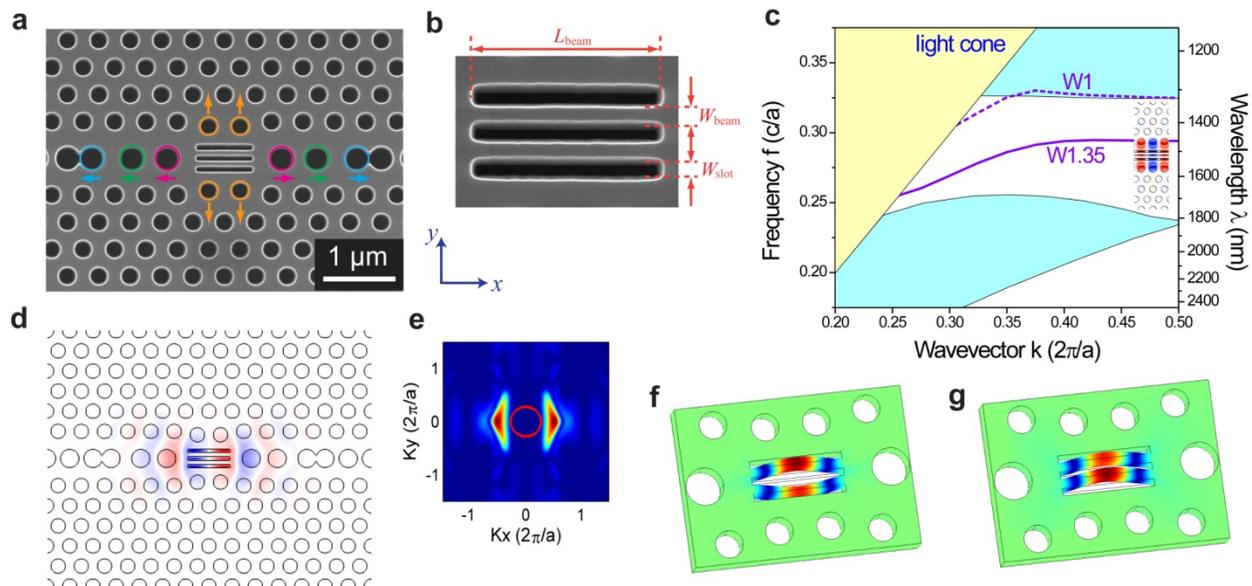

Figure 1. (a) Scanning electron micrograph of the nanomechanical double-beam resonator in a PhC nanocavity. The color coded holes around the double beams are shifted (outward defined as positive shift). (b) Close-up view of the double beams. (c) PhC band diagram of a W1.35 waveguide and a W1 waveguide, both with triple slots and formed in a 220-nm Si layer. The three 60-nm slots are separated by two 80-nm beams in between. The continuum of PhC slab modes is indicated by the cyan areas. A single guided band of the W1.35 waveguide (violet solid line) is present at the middle of the PhC bandgap, while the corresponding band of the W1 waveguide (violet dash line) is buried in the upper continuum of the PhC slab modes. (d) Simulated TE-like electrical field component $E_y$ of the resonant mode ($\lambda_0 = 1541.7$ nm) of the optimized double beam in PhC nanocavity. The profile indicates that the optical mode closely

matches the mechanical mode volume. (e) Two-dimensional Fourier transform of the electric field. The red circle at the center defines the leaky region, i.e., the inside of the light cone shown in (c). (f, g) Simulated mechanical mode of the double beams, with differential (f) and common motion (g).

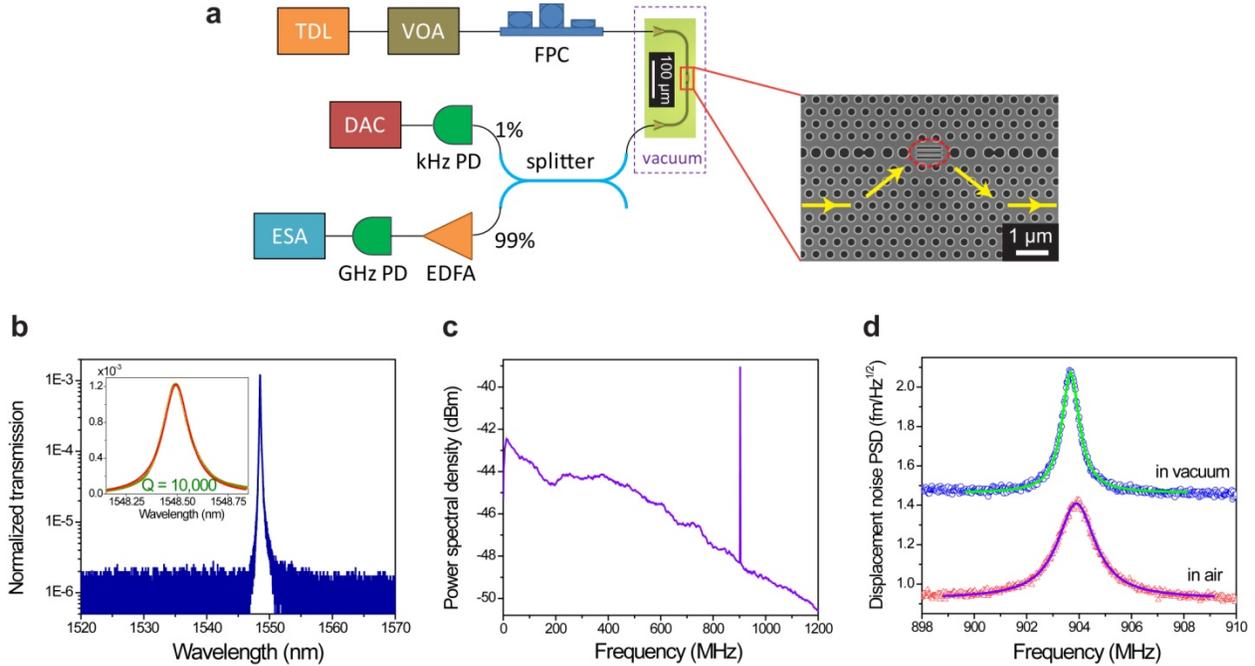

Figure 2. (a) Experimental setup. TDL, tunable diode laser; VOA, variable optical attenuator; FPC, fiber polarization controller; PD, photodetector; DAC, data acquisition card; EDFA, erbium-doped fiber amplifier; ESA, electrical spectrum analyzer. (b) Normalized optical transmission spectrum, showing a single optical resonance with a loaded optical $Q$ of 10,000. (c) RF power spectrum density of the light transmitted through an $L_{beam}$ = 785 nm, $W_{beam}$ = 80 nm device, exhibiting optical transduction of a single mechanical mode in the entire measurement range. (d) Zoomed-in RF spectra of the same device exhibiting mechanical $Q$ values of 1230 in vacuum and 580 in air measured with 1.6 and 2.5 mW optical power in the feeding waveguide respectively. A detector-noise-limited displacement sensitivity of 0.94 fm/Hz$^{1/2}$ is obtained for the measurement in air.

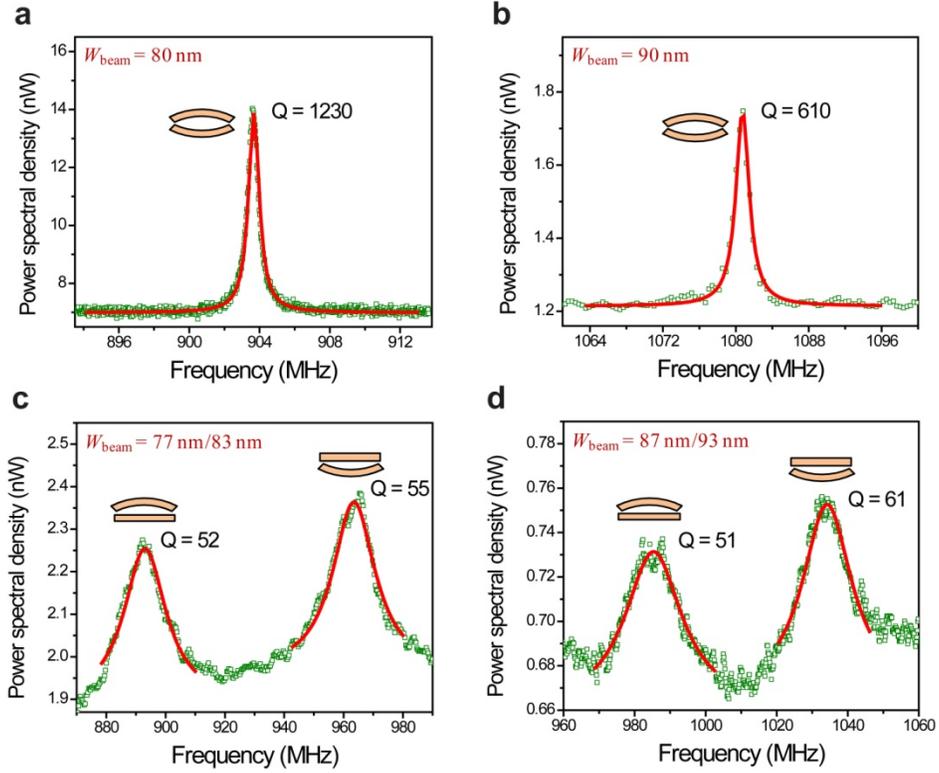

Figure 3. (a) RF power spectrum of a symmetric-design $W_{beam}$ = 80 nm device showing a single high-$Q$ peak at 903.6 MHz. (b) RF power spectrum of a symmetric-design $W_{beam}$ = 90 nm device showing a single high-$Q$ peak at 1.081 GHz. (c) RF power spectrum of an asymmetric-design $W_{beam}$ = 77 nm/83 nm device showing double low-$Q$ peaks. (d) RF power spectrum of an asymmetric-design $W_{beam}$ = 87 nm/93 nm device showing double low-$Q$ peaks. All the devices have a beam length $L_{beam}$ of 785 nm.

# Supporting Information for "Femtogram Doubly Clamped Nanomechanical Resonators Embedded in a High-$Q$ Two-Dimensional Photonic Crystal Nanocavity"

Xiankai Sun,[1] Jiangjun Zheng,[2] Menno Poot,[1] Chee Wei Wong,[2] and Hong X. Tang[1,*]

[1]Department of Electrical Engineering, Yale University, New Haven, Connecticut 06511, United States
[2]Optical Nanostructures Laboratory, Columbia University, New York, New York 10027, United States
*Corresponding author: hong.tang@yale.edu

## A. Derivation of Eigenmodes of Coupled Oscillator System

In this section we present the detailed derivation of the solutions to the coupled equations (1) and (2) of the main text. We also derive the modal frequency and damping rate of the coupled modes in several typical situations.

As illustrated in Figure S1, two mechanical beam oscillators possessing different masses ($m_1$ and $m_2$) and different elastic constants ($k_1$ and $k_2$) are coupled to each other with both dissipative and dispersive coupling. The equations of motion that govern this coupled oscillator system are[1,2]

$$m_1 \ddot{u}_1 + m_1 \gamma_a \dot{u}_1 + \gamma_c (m_1 \dot{u}_1 + m_2 \dot{u}_2) + k_1 u_1 + k_c (u_1 - u_2) = F_1, \tag{S1}$$

$$m_2 \ddot{u}_2 + m_2 \gamma_a \dot{u}_2 + \gamma_c (m_1 \dot{u}_1 + m_2 \dot{u}_2) + k_2 u_2 + k_c (u_2 - u_1) = F_2. \tag{S2}$$

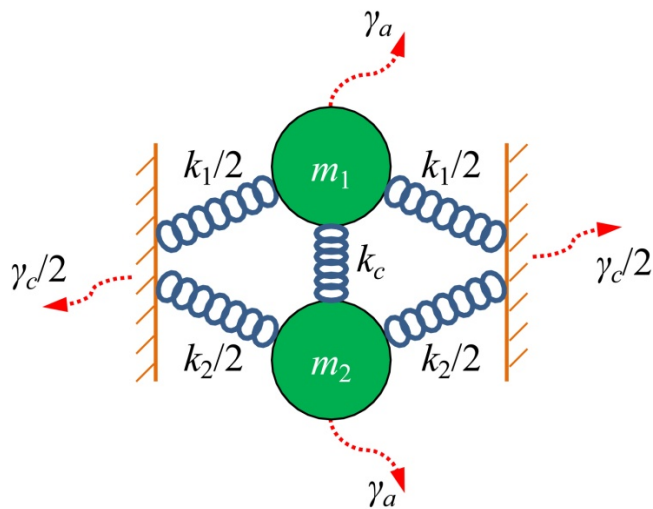

Figure S1. Model of two coupled mechanical beam oscillators.

In the above equations, $\gamma_a$ is the damping rate for all coupling-independent dissipation channels, such as material loss, air damping, and the coupling-independent portion of clamping loss. The $\gamma_c$ terms denote the additional clamping loss, which is assumed to be proportional to the velocity of the center of mass of the coupled oscillator system. $k_c$ is the dispersive coupling coefficient of the two oscillators. Finally, $F_1$ and $F_2$ are the thermal Langevin forces acting on the oscillators, which are correlated due to the coupling terms. Note that, in principle, $\gamma_c$ can be positive or negative; the latter would indicate that the *differential* mode has a higher damping. Since the experimental data confirm the more intuitive picture where the *common* mode has the highest damping, we will assume that $\gamma_c$ is positive in the following discussion.

To find the eigenmodes of the coupled oscillator system, $F_1$ and $F_2$ are set to zero and harmonic motion is assumed for the two displacements: $u_1 = Ae^{i\Omega t}$ and $u_2 = Be^{i\Omega t}$. Then by defining

$$\Omega_{1,2}^2 = \frac{k_{1,2} + k_c}{m_{1,2}},$$

(S1) and (S2) are expressed in a matrix form

$$\begin{pmatrix} -\Omega^2 + i\Omega(\gamma_a + \gamma_c) + \Omega_1^2 & i\Omega\gamma_c \frac{m_2}{m_1} - \frac{k_c}{m_1} \\ i\Omega\gamma_c \frac{m_1}{m_2} - \frac{k_c}{m_2} & -\Omega^2 + i\Omega(\gamma_a + \gamma_c) + \Omega_2^2 \end{pmatrix} \begin{pmatrix} A \\ B \end{pmatrix} = 0, \quad (S3)$$

The values of $\Omega$ where the matrix is not invertible correspond to the eigenmodes; the real and imaginary part of $\Omega$ are the eigenfrequency and (half of) the damping rate of the mode, respectively.

To simplify the analysis, the following quantities are introduced: the mean frequency $\bar{\Omega} = \frac{\Omega_1 + \Omega_2}{2}$, the detuning $s = \Omega - \bar{\Omega}$, the frequency asymmetry $\delta = \frac{\Omega_1 - \Omega_2}{2}$, and the coupling rate $v_{c1,2} = \frac{k_c}{m_{1,2}\bar{\Omega}}$. Then, with the assumptions $|\delta|, |s|, \gamma_a, \gamma_c \ll \Omega, \bar{\Omega}$, the approximations $\Omega_{1,2}^2 = (\bar{\Omega} \pm \delta)^2 \approx \bar{\Omega}^2 \pm 2\bar{\Omega}\delta$ and $\Omega^2 = (\bar{\Omega} + s)^2 \approx \bar{\Omega}^2 + 2\bar{\Omega}s$ hold, which correspond to the assumption of a Lorentzian response of the eigenmodes. As a result, (S3) is simplified into

$$\begin{pmatrix} -2\bar{\Omega}s + i\bar{\Omega}(\gamma_a + \gamma_c) + 2\bar{\Omega}\delta & i\bar{\Omega}\gamma_c \frac{m_2}{m_1} - \bar{\Omega}v_{c1} \\ i\bar{\Omega}\gamma_c \frac{m_1}{m_2} - \bar{\Omega}v_{c2} & -2\bar{\Omega}s + i\bar{\Omega}(\gamma_a + \gamma_c) - 2\bar{\Omega}\delta \end{pmatrix} \begin{pmatrix} A \\ B \end{pmatrix} = 0. \quad (S4)$$

By setting the determinant of the coefficient matrix to zero, we obtain the characteristic equation

$$s^2 - is(\gamma_a + \gamma_c) - \left(\frac{\gamma_a + \gamma_c}{2}\right)^2 - \delta^2 - \frac{1}{4}(i\gamma_c - v_{c1})(i\gamma_c - v_{c2}) = 0, \quad (S5)$$

with the solutions given by

$$s = i\frac{\gamma_a + \gamma_c}{2} \pm \left[\delta^2 + \frac{1}{4}(i\gamma_c - v_{c1})(i\gamma_c - v_{c2})\right]^{1/2}, \tag{S6}$$

leading to the eigenvalues of (S3)

$$\Omega = \bar{\Omega} + s = \frac{\Omega_1 + \Omega_2}{2} + i\frac{\gamma_a + \gamma_c}{2} \pm \left[\delta^2 + \frac{1}{4}(i\gamma_c - v_{c1})(i\gamma_c - v_{c2})\right]^{1/2}. \tag{S7}$$

The modal frequencies and the corresponding damping rates can be simplified in several typical cases that are of great interest:

1) In the absence of dissipative and dispersive coupling channels, i.e., $k_c = \gamma_c = 0$ ($v_{c1} = v_{c2} = 0$),

$$\Omega = \frac{\Omega_1 + \Omega_2}{2} + i\frac{\gamma_a}{2} \pm |\delta| = \Omega_{1,2} + i\frac{\gamma_a}{2},$$

which recovers the solutions of the individual uncoupled oscillators.

2) If two identical oscillators are coupled, $\delta = 0$, $v_{c1} = v_{c2} \equiv v_c$,

$$\Omega = \frac{\Omega_1 + \Omega_2}{2} + i\frac{\gamma_a + \gamma_c}{2} \pm \frac{i\gamma_c - v_c}{2} = \begin{cases} \frac{\Omega_1 + \Omega_2}{2} - \frac{v_c}{2} + i\left(\frac{\gamma_a}{2} + \gamma_c\right), & \text{where } (A\ B)^T = (1\ 1)^T, \\ \frac{\Omega_1 + \Omega_2}{2} + \frac{v_c}{2} + i\frac{\gamma_a}{2}, & \text{where } (A\ B)^T = (1\ -1)^T, \end{cases}$$

which describes a common mode with a higher damping rate $\gamma_a + 2\gamma_c$, and a differential mode with a lower damping rate $\gamma_a$.

3) If two nonidentical but similar oscillators are coupled with the degree of asymmetry represented by $|\delta|$, $\delta^2 \ll \frac{1}{4}(i\gamma_c - v_{c1})(i\gamma_c - v_{c2}) \approx \frac{1}{4}(i\gamma_c - v_c)^2$,

$$\Omega \approx \frac{\Omega_1 + \Omega_2}{2} + i\frac{\gamma_a + \gamma_c}{2} \pm \frac{i\gamma_c - v_c}{2} \mp \frac{\delta^2(i\gamma_c + v_c)}{\gamma_c^2 + v_c^2}$$

$$= \begin{cases} \frac{\Omega_1 + \Omega_2}{2} - \frac{v_c}{2}\left(1 + \frac{2\delta^2}{\gamma_c^2 + v_c^2}\right) + i\left(\frac{\gamma_a}{2} + \gamma_c\left(1 - \frac{\delta^2}{\gamma_c^2 + v_c^2}\right)\right), & \text{where } (A\ B)^T \approx (1\ 1)^T, \\ \frac{\Omega_1 + \Omega_2}{2} + \frac{v_c}{2}\left(1 + \frac{2\delta^2}{\gamma_c^2 + v_c^2}\right) + i\left(\frac{\gamma_a}{2} + \gamma_c \frac{\delta^2}{\gamma_c^2 + v_c^2}\right), & \text{where } (A\ B)^T \approx (1\ -1)^T, \end{cases}$$

which describes a common mode with a modified damping rate $\gamma_a + 2\gamma_c\left(1 - \frac{\delta^2}{\gamma_c^2 + v_c^2}\right)$, and a differential mode with a modified damping rate $\gamma_a + 2\gamma_c \frac{\delta^2}{\gamma_c^2 + v_c^2}$.

4) If two considerably dissimilar oscillators are coupled, $\delta^2 \gg \frac{1}{4}(i\gamma_c - v_{c1})(i\gamma_c - v_{c2})$,

$$\Omega \approx \frac{\Omega_1 + \Omega_2}{2} + i\frac{\gamma_a + \gamma_c}{2} \pm |\delta| = \begin{cases} \Omega_1 + i\frac{\gamma_a + \gamma_c}{2}, & \text{where } (A \quad B)^T \approx (1 \quad 0)^T, \\ \Omega_2 + i\frac{\gamma_a + \gamma_c}{2}, & \text{where } (A \quad B)^T \approx (0 \quad 1)^T, \end{cases}$$

which describes that the coupled modes possess frequencies of the uncoupled oscillators with equal damping rates $\gamma_a + \gamma_c$.

Note that case 1 should not be confused with case 4. In case 1 the two oscillators do not have coupling at all because all the coupling channels are turned off, while in case 4 although two considerably dissimilar oscillators have negligible mutual coupling, they still individually couple to the environment via an additional dissipative channel. The damping rates in case 4 are thus higher than those in case 1. Based on (S7), the evolution of damping rates $2\text{Im}(\Omega)$ for the in-phase and anti-phase mode as a function of the asymmetric parameter $|\delta|$ is plotted in Figure S2 for different ratios of $v_c$ to $\gamma_c$ (where $v_{c1} = v_{c2} \equiv v_c$ has been assumed).

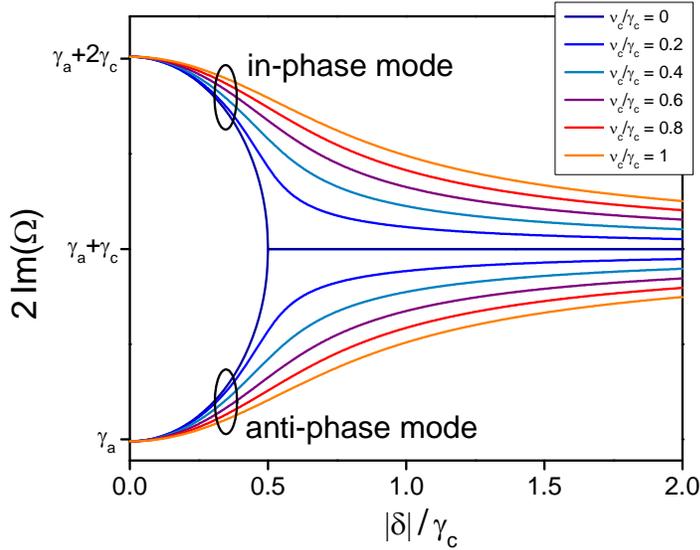

Figure S2. Evolution of the damping rates of the eigenmodes of the coupled oscillator system for $v_{c1} = v_{c2} \equiv v_c$.

### B. Stress and Displacement Fields of the Eigenmodes

The stress and displacement fields of the eigenmodes of the symmetric double beams in photonic crystal cavity are obtained using finite-element simulation with a perfectly-matched-layer boundary condition.[3] When comparing the field intensities at the far-field (close to the outer boundary) shown in Figure S3, it is clear that the dipole-like radiation caused by the common motion (a) is orders of magnitude stronger than the quadruple-like radiation caused by the differential mode (b). This indicates that the common motion induces much stronger radiation to the substrate, or equivalently, the vibrations of the common mode inside the mechanical resonator have much larger tunneling probability to outside phonon continuum.[4] The associated dissipation rate (1/$Q$) can be obtained by an overlap integral involving the scattering mode and the resonator mode. Since the $Q$ value is inversely related to the stress at the far-field,[4] it can be inferred that the common mode will thus possess a substantially lower $Q$ value than the differential mode.

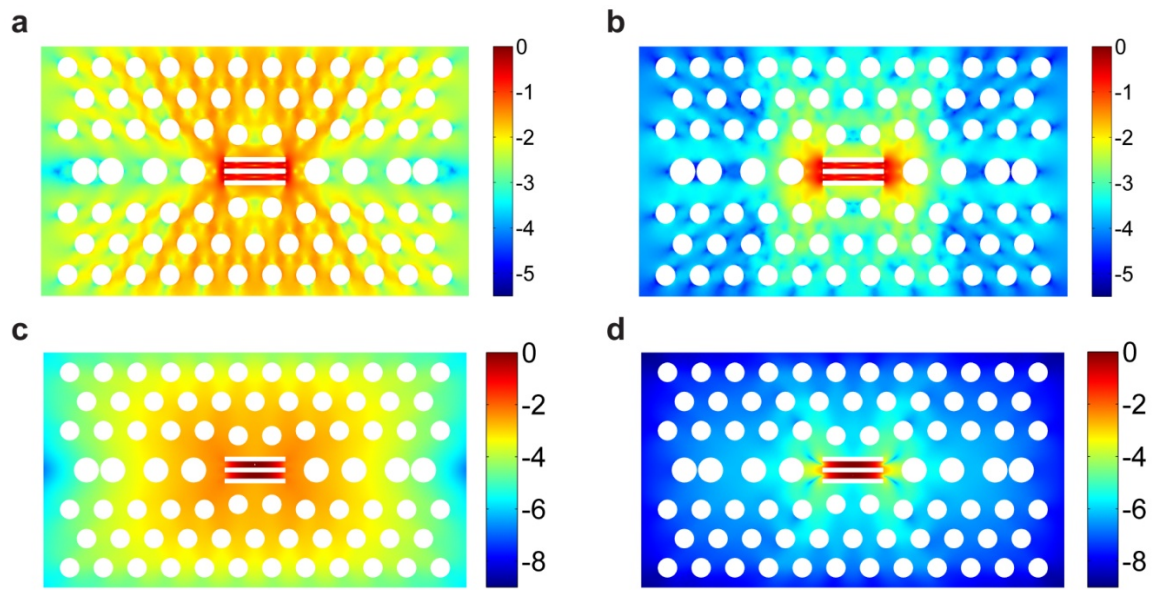

Figure S3. (a, b) Normalized von Mises stress field (in log scale) of the common mode (a) and differential mode (b). Note the dipole-like field of the common mode (a) is orders of magnitude stronger than the quadruple-like field of the differential mode (b) at the far-field, which indicates that the common mode experiences much stronger radiation leakage to the substrate and thus possesses a substantially lower $Q$ value. (c, d) Displacement field intensity $|\mathbf{U}|^2/\max(|\mathbf{U}|^2)$ (in log scale) of the common mode (c) and differential mode (d). Orders of magnitude difference at the far-field is also distinct.